\documentclass[prb,final,twocolumn,superscriptaddress,showpacs]{revtex4}
\usepackage{graphicx}
\usepackage{color}

\begin{document}

\title{$\omega$/T scaling of the optical conductivity in strongly correlated layered cobalt oxide}
\author{P. Limelette}
\affiliation{Universit\'e Fran\c{c}ois Rabelais, Laboratoire GREMAN, UMR 7347 CNRS, Parc de Grandmont, 37200 Tours, France}
\author{V. Ta Phuoc}
\affiliation{Universit\'e Fran\c{c}ois Rabelais, Laboratoire GREMAN, UMR 7347 CNRS, Parc de Grandmont, 37200 Tours, France}
\author{F. Gervais}
\affiliation{Universit\'e Fran\c{c}ois Rabelais, Laboratoire GREMAN, UMR 7347 CNRS, Parc de Grandmont, 37200 Tours, France}
\author{R. Fr\'esard}
\affiliation{Laboratoire CRISMAT, UMR 6508 CNRS-ENSICAEN, 6, Boulevard du Mar\'echal Juin, 14050 CAEN Cedex, France}
\begin{abstract}
\vspace{0.3cm}
We report infrared spectroscopic properties of the strongly correlated layered
cobalt oxide [BiBa$_{0.66}$K$_{0.36}$O$_2$]CoO$_2$. These measurements
performed on single crystals allow us to determine the optical conductivity as a function of temperature.
In addition to a large temperature dependent transfer of spectral weight,
an unconventional low energy mode is found. We show that both its frequency
and  damping scale as the temperature itself. In fact, a basic
analysis demonstrates that this mode fully scales onto a function of
$\omega$/T up to room temperature. This behavior suggests low energy excitations of non-Fermi liquid
type originating from quantum criticality. 
\end{abstract}


\maketitle
Beyond the successful Fermi liquid paradigm, the non-Fermi liquid states of matter are nowadays the subject of intense both theoretical \cite{Jain2009,Sachdev2010,Phillips2011} and experimental \cite{Stewart2001,Lohneysen2007} investigations.
These states can in particular be reached in the vicinity of a so-called quantum critical point (QCP) which is located at zero temperature and results from competing interactions tuned by an appropriate non-thermal control parameter, such as pressure, doping or magnetic field \cite{Sachdev2011}.
As quite recently raised by A. Kopp and S. Chakravarty for instance \cite{Chakravarty}, "a fundamental question is how high in temperature can the effects of quantum criticality persist" and if "physical observables could be described in terms of universal scaling functions originating from the QCPs" over an extended temperature range.
While they demonstrate that the temperature can be surprisingly high, experimental evidences remain yet to be given to support these 
theoretical predictions.
Also, famous examples of non-Fermi liquid behaviors near QCPs are provided by
heavy fermion metals \cite{Custers2003,Lohneysen2007,Schroder2000} which have
early on emerged as prototypical materials to investigate such unconventional
properties, and also by some transition metal oxides
\cite{Gegenwart2006,Vandermarel2003}, including the layered cobalt oxide [BiBa$_{0.66}$K$_{0.36}$O$_2$]CoO$_2$ \cite{Limelette2010-a}.
More specifically, the latter compound belongs to a family of doped Mott insulators 
\cite{Imada1998,Georges1996,Devaulx2007} which has revealed striking properties \cite{Bobroff2007,Schulze2008,Nicolaou2010} such as enhanced room temperature thermopower \cite{Terasaki1997,Wang2003} or large negative magnetoresistance \cite{Limelette2008-a}.
Interestingly, the so-called giant electron-electron scattering inferred in Na$_{0.7}$CoO$_2$ \cite{Li2004} has already led to conjecture a possible influence of a magnetic QCP in a qualitative agreement with the density functional calculations predicting weak itinerant ferromagnetic state competing with weak itinerant antiferromagnetic state \cite{Singh2007}.

In this context, recent susceptibility \cite{Limelette2010-a} and specific
heat measurements \cite{Limelette2010-b} have given good evidence that most of
the electronic properties observed in the metallic cobalt oxide
[BiBa$_{0.66}$K$_{0.36}$O$_2$]CoO$_2$ display unconventional behaviors which
seem to originate from a magnetic field induced QCP. In fact, the investigated susceptibility $\chi$ has revealed a scaling behavior \cite{Limelette2010-a} with both the temperature $T$ and the magnetic field $B$ ranging from a high-$T$ non-Fermi liquid down to a low-$T$ Fermi liquid.
In the latter Fermi liquid regime, the divergent behavior of the Pauli
susceptibility follows a power law dependence $\chi \propto$ b$^{-0.6}$. Here
$b=B-B_C$ measures the distance from the QCP and the critical magnetic field
$B_C \approx 0.176\ T$.
Furthermore, specific heat measurements allowed to interprete the enhancement
of the Pauli susceptibility as a result of efficient ferromagnetic
fluctuations by analyzing the Wilson ratio \cite{Limelette2010-b}. 

\begin{figure}[htb]
\centerline{\includegraphics[width=0.95\hsize]{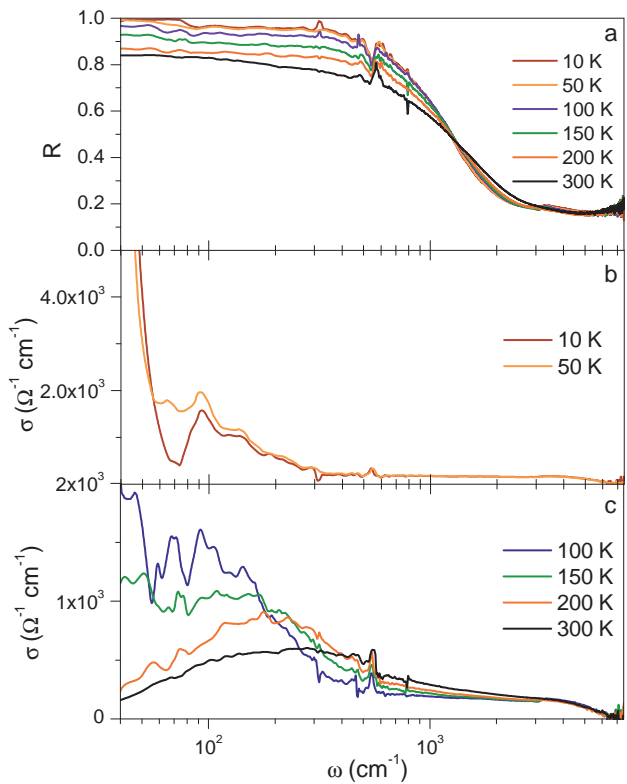}}
\caption{(Color online) Frequency dependence of the single crystal
  reflectivity (a), and optical conductivity (b) and (c), as functions of
  temperature from 10~K up to 300~K. $\sigma(\omega)$ has been obtained with
  $\epsilon_{\infty} \approx$ 7.} 
\label{fig-1}
\end{figure}

The purpose of this Letter is to acquire more insight into the nature of the
electronic excitations of 
the cobalt oxide [BiBa$_{0.66}$K$_{0.36}$O$_2$]CoO$_2$ by investigating
its infrared spectroscopic properties, and especially its scaling behavior.  
Therefore, reflectivity measurements have been performed with a Fourier transformed infrared spectrometer Bruker IFS 66v/S spanning the frequency interval from 40 cm$^{-1}$ up to 8000 cm$^{-1}$ and, as a function of temperature, down to 10~K.
It is worth mentioning that normal incidence radiation has been used in order to measure the in-plane single crystal reflectivity.
After the initial measurement, the sample was coated in situ with a gold film and re-measured at each temperature.
These additional data were used as reference mirrors to calculate the
reflectivity in order to take into account light scattering on the surface of
the sample. 
Let us also note that the reported results have been measured with
several single crystals thereby ensuring the reproducibility of these
investigations. We emphasize that our results also agree with previous reflectivity measurements which have not led to a determination of the optical conductivity \cite{Dong2008}, in contrast to the analysis reported in this Letter.
Figure \ref{fig-1} displays the frequency dependence of the reflectivity within the experimental range as a function of the temperature from 300~K down to 10~K.
The latter already illustrates an unusual temperature dependence by showing a
huge transfer of spectral weight  from high to low energies in agreement with previous reflectivity measurements.
Here, we propose an analysis of these data by determining the optical
conductivity and by investigating its unconventional temperature
dependence. Indeed, the reflectivity may be related to the complex dielectric
constant $\epsilon(\omega)$ following the Fresnel relation 
$R(\omega)= \left| (1-\sqrt{\epsilon(\omega)})/(1+\sqrt{\epsilon(\omega)}
\right|^2 $. Then, the real part of the optical conductivity
$\sigma(\omega)$ follows from the imaginary part of the dielectric constant
as $\sigma(\omega)= 2 \pi c \omega \epsilon''(\omega)$, $\omega$
being here the frequency in the unit of the wavelength inverse and $c$ the speed of light.
Though the complex dielectric constant can be obtained through the standard
Kramers-Kronig transformation we rather proceed with the usual fitting routine with Drude-Lorentz oscillators as:
$
\epsilon(\omega)= \epsilon_{\infty}+ \epsilon_0 \sum_i \frac{\omega_{p,i}^2}{\omega_{0,i}^2-\omega^2-i \gamma_i \omega} 
$. 
Here, $\epsilon_{\infty}$ is related to an effective contribution of all the high frequency oscillators out of the experimental range, while $\omega_{p,i}$, $\omega_{0,i}$ and $\gamma_i$ are the effective plasma frequency, the frequency, and the
damping of the i$^{th}$ mode, respectively.
Note that a Drude term may be considered to take free charge carriers into account by vanishing the frequency $\omega_{0,i}$.%

In order to determine the optical conductivity, we used a variational routine
with a causal dielectric function, namely  Kramers-Kronig consistent, which provides the best match to the experimental reflectivity by using typically 500 oscillators.
It must be emphasized that according to this procedure, these oscillators have not necessarily a physical microscopic origin which makes this routine closer to a fitting procedure rather than to a modeling one.
Yet, such a procedure is equivalent to the standard Kramers-Kronig analysis \cite{Kuzmenko2005}.
Figure \ref{fig-1} shows the variations of the optical conductivity determined using the previous routine within the frequency range from 40 cm$^{-1}$ up to 8000 cm$^{-1}$.
In particular, the low frequency optical conductivity determined at 300 K is of the order of 200 $\Omega^{-1} cm^{-1}$ in agreement with the previously reported DC value 250 $\Omega^{-1} cm^{-1}$ \cite{Dong2008}.
A strong temperature dependence is also observed with spectral weight redistribution over an anomalously large frequency interval of the order of the experimental range.
We note a transfer of spectral weight from high energies down to low energies occurs when the temperature is decreased, in order to form a Drude-like peak at low frequency below 50~K.\\
\begin{figure}[htb]
\centerline{\includegraphics[width=0.95\hsize]{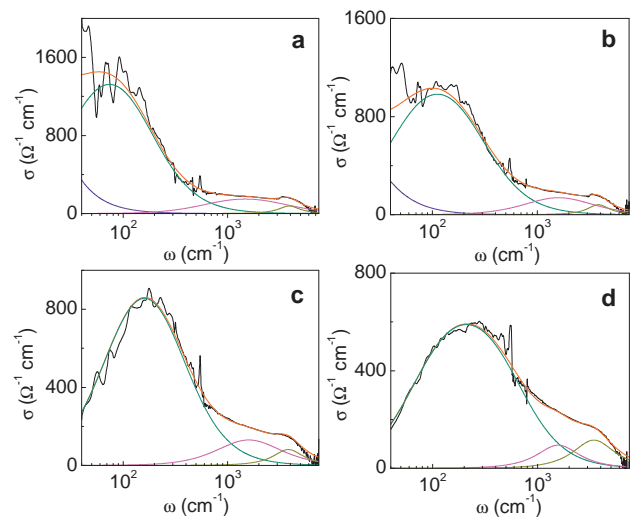}}
\caption{(Color online) Results of the modeling of the global shape of the reflectivity with the variation of the optical conductivity as a function of frequency at 100~K (a), 150~K (b), 200~K (c) and 300~K (d).
Three Drude-Lorentz oscillators have been used with the frequencies $\omega_{0,i} \approx$ k$_B$T/hc, 1400, 4000 cm$^{-1}$ and a very weak Drude-like contribution at 100~K and 150~K.}
\label{fig-fit-sigma}
\end{figure}

\begin{figure}[htb]
\centerline{\includegraphics[width=0.95\hsize]{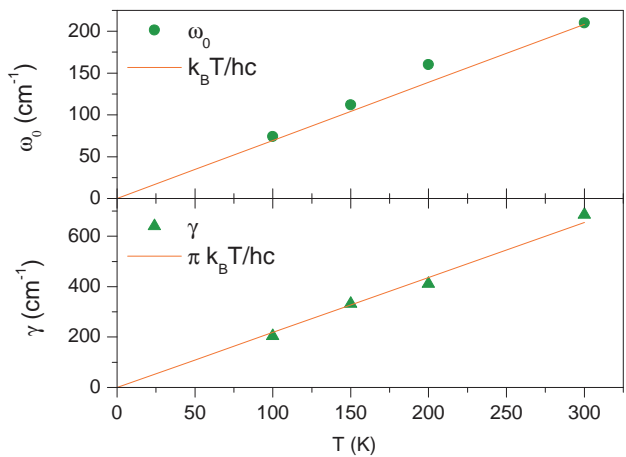}}
\caption{(Color online) Temperature dependence of the frequency $\omega_0$ and the damping $\gamma$ for the low frequency oscillator.}
\label{fig-qc-freq}
\end{figure}

From a qualitative point of view, let us mention that such a spectral
redistribution has already been observed in materials near the Mott metal to insulator transition where low energy electronic excitations strongly couple to the high energy ones \cite{Georges1996}.
Therefore, the behavior of the optical conductivity is likely due to strong
electronic correlations as exemplified by the narrow width of the Drude-like
peak which suggests renormalized quasiparticle energies. 
Interestingly, Fig.~\ref{fig-1}c exhibits an unusual feature above 100~K as a broad maximum with a frequency which increases with temperature and seems to amazingly coincide with $k_B$T, $k_B$ being the Boltzmann constant.\\
\indent Beyond the previous determination of the optical conductivity, a modeling procedure has been performed 
in order to analyze quantitatively the aforementioned unusual maximum. 
Thus, a minimal number of Drude-Lorentz oscillators has been used to account for the global shape of the reflectivity  \cite{remrk}.
As shown in Fig.~\ref{fig-fit-sigma}, three Drude-Lorentz oscillators have been identified to basically account for the temperature dependence of the optical conductivity as a function of frequency, with a very weak Drude-like contribution below 150~K.
These figures all reveal an important contribution, namely with a sizeable spectral weight, originating from a low frequency oscillator.
They also already demonstrate that the frequency of the latter mode $\omega_0$
increases continuously with the temperature, in a linear fashion.
This amazing behavior is indeed checked by plotting in Fig.~\ref{fig-qc-freq} the temperature dependence of this frequency.
Moreover, it appears that not only $\omega_0$ but also the damping $\gamma$ are directly proportional to the temperature as written below, with their contribution to the conductivity $\sigma^*$.
\begin{equation}
\omega_0 = \frac{k_B T}{h c} = \frac{ \gamma }{ \pi}   
\mbox{\ \ \ and  \ \ \ }  \sigma^* =  \frac{2 \pi c  \epsilon_0 \gamma \omega^2 \omega_{p}^2 }{(\omega_{0}^2-\omega^2)^2+ \gamma^2 \omega^2} 
\label{eq-w_0-chi}
\end{equation}
In addition, it is worth mentioning that the temperature dependence of $\gamma$ is very close to $\pi k_B T/hc$, as demonstrated in Fig.~\ref{fig-qc-freq}.

\begin{figure}[htb]
\centerline{\includegraphics[width=0.95\hsize]{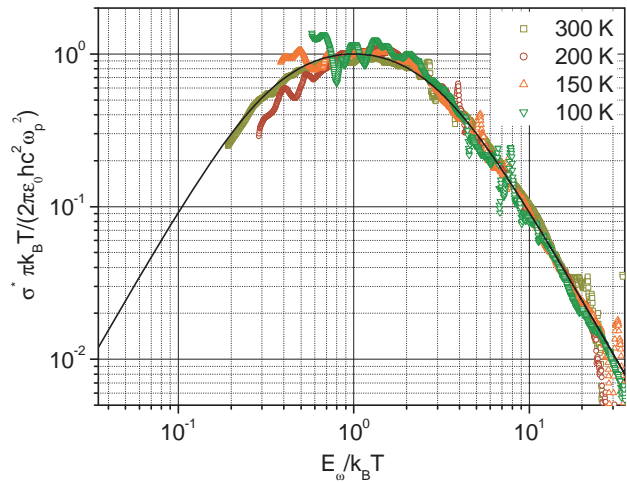}}
\caption{(Color online) Scaling of the unconventional contribution to the optical conductivity as a function of the dimensionless ratio $E_{\omega}/k_B T$ for T=100, 150, 200, 300~K as explained in the text.
Note that this contribution is here plotted as the dimensionless quantity $\sigma^* \pi k_B T/(2\pi \epsilon_0 h c^2 \omega_p^2 )$.
The full line represents the right hand side of Eq.~(\ref{eq-sigma-scaling}).
}
\label{fig-optic-scaling}
\end{figure}

While not shown here, it should be noted that the effective plasma frequency of this oscillator seems to exhibit a moderate and rather linear increase with temperature as $\omega_p \approx 3653+4.5 T$.
Nevertheless, one cannot firmly ascribe a linear power law dependence because of the weakness of the latter variation.
%
%
On the other hand, according to the linear temperature dependences of both
$\omega_0$ and $\gamma$ in Eq.~(\ref{eq-w_0-chi}), the contribution $\sigma^*$
actually appears to only depend on the dimensionless ratio between the frequency $\omega$ and the temperature following Eq.~(\ref{eq-sigma-scaling}), with the photon energy E$_{\omega}$= hc $\omega$ and $x=E_{\omega}/k_B T$.
\begin{equation}
\frac{ k_B T }{ 2 h c^2  \epsilon_0 \omega_{p}^2 } \sigma^* =  \frac{\pi^2 x^2 }{\pi^2 x^2 + (1-x^2)^2} 
\label{eq-sigma-scaling}
\end{equation}
\indent As a result, one should be able to highlight such a contribution by
subtracting those of the other tow oscillators displayed in Fig.~\ref{fig-fit-sigma} and corresponding to the frequencies $\omega_{0,i}$ around 1400 cm$^{-1}$ and 4000 cm$^{-1}$.
By using the values of the effective plasma frequency as given by the aforementioned approximate linear relation and the parameters of the two high energy Drude-Lorentz oscillators, including the small Drude part at 100~K and 150~K shown in Fig.~\ref{fig-fit-sigma}, the residual low energy contribution can thus be selected and plotted as a dimensionless quantity as a function of the dimensionless ratio $E_{\omega}/k_B T$.
As a strong check of the overall analysis, Fig.~\ref{fig-optic-scaling} shows that the data related to $\sigma^*$ nicely collapse onto a single curve over nearly two decades in full agreement with Eq.~(\ref{eq-sigma-scaling}).
The normalized maximum is in particular recovered for $E_{\omega}/k_B T$ = 1, which firmly checks the unconventional linear  temperature dependences of both $\omega_0$ and $\gamma$ as displayed in Fig.~\ref{fig-qc-freq}.

Since neither the spectral weight nor the scaling properties as a function of
$\omega/T$ can easily be ascribed to phonons \cite{Eagles1995} they suggest
that the relevant energy scale of these likely electronic excitations is given by the temperature.
%
%
Whereas the analyzed oscillator does not contribute to the zero frequency resistivity at finite temperatures, 
this kind of scaling shares similarities with some properties measured in the high temperature cuprates superconductors.
Indeed, Van Der Marel et al gave good evidence that the optical conductivity of some cuprates follows
scaling laws as $\sigma(\omega)$=T$^{-\mu}$g($\omega/T$) \cite{Vandermarel2003}.
In fact two different scaling functions were found: One in the far infrared
for $\omega/T < 1.5$ with $\mu = 1 $, and one in an intermediate frequency
range (yet with  $\omega/T > 3$) with $\mu = 0.5$, stressing the role of the
temperature as a crossover frequency put forward by Damle and
Sachdev \cite{Damle}. One should also mention that the scaling form $\sigma(\omega)$T = g($\omega/T$) is automatically fulfilled if the only contribution to the optical conductivity is Drude-like, namely with $\omega_0$=0 in Eq.~(\ref{eq-w_0-chi}), with a damping varying linearly with temperature $\gamma \propto T$.
In this context, the previous scaling form implies, by considering the zero
frequency limit, a resistivity which increases linearly with the temperature (with here g(0)$\neq$ 0) as experimentally observed.
Therefore, the origin of such a scaling form appears as a central issue since
it is directly related to the enigmatic transport in the so-called normal
state in cuprates. 
This is commonly interpreted in terms of quantum critical fluctuations originating from a QCP ($T=0$). 
The distance from such a QCP being given by k$_B$T, the temperature is then the only relevant energy scale governing the electronic excitations in the quantum critical regime.
In addition, fluctuation energy is inversely proportional to the correlation time $\tau$ and is also related to the correlation length $\xi$ as $\omega \propto \tau^{-1} \propto \xi^{-z}$, with the dynamical exponent z.
In the framework where the quantum critical behavior 
is accounted for by the one-parameter scaling hypothesis, Phillips and Chamon \cite{Phillips2005} have
demonstrated that charge conservation implies the following general
scaling form for the optical conductivity: 
\begin{equation}
\sigma (\omega,T) =  \frac{e^2}{h} \xi^{2-d} g \left( \omega \tau \right)~~,
\label{eq-sigma-scaling-th-1}
\end{equation}
with a dimensionless function $g(\omega
\tau)$, the elementary charge e and the spatial dimension d. In
Eq.~(\ref{eq-sigma-scaling-th-1}) the optical conductivity is expressed in
units of the quantum of conductance $e^2/h$. Further, one may infer the
scaling form Eq.~(\ref{eq-sigma-scaling-th-2}) as a function of both the
temperature and the frequency as: 
\begin{equation}
\sigma (\omega,T) =  \frac{e^2}{h} \left( \frac{k_B T}{h c_0} \right)^{(d-2)/z} g\left( \frac{\omega }{T} \right) ~~,
\label{eq-sigma-scaling-th-2}
\end{equation}
%
with $c_0$ a non-universal velocity for $z=1$ \cite{Damle}. It results that in a three dimensional system, even anisotropic as in the case of cuprates or in the case of this layered cobalt oxide, the scaling $\sigma (\omega,T) \propto $ T$^{-1}$g($\omega $/T) implies an unphysical negative dynamical exponent z=-1 and thus, an inconsistency between the scaling form Eq.~(\ref{eq-sigma-scaling}) and the generic equation~(\ref{eq-sigma-scaling-th-2}).
Furthermore, the observation of such a scaling behavior over a vast
temperature range and up to 300\ K invalids most of the other  possible
interpretations, in particular in terms of a classical phase
transition. Indeed, its physics is driven by thermal fluctuations which are
only critical in a small temperature range, basically around the transition
temperature, that vanishes in a quantum critical system. On the contrary,
it has been predicted theoretically that quantum
criticality can extend up to surprisingly large temperatures, that can even
exceed  room temperature, because they are essentially bounded by the bare energy scale \cite{Chakravarty}.
Therefore, while the scaling behavior here observed over two decades can be taken as an evidence for the persistence of a quantum criticality, the scaling function itself points towards its unconventional nature.
This could require a description beyond the one-parameter paradigm \cite{Phillips2005} as already supported by the previously reported susceptibility measurements \cite{Limelette2010-a}. 
Let us finally mention that our method based on the extraction of the quantum critical contribution, 
could be put under further test in other strongly correlated systems as in
some cuprates \cite{lupi2000}, in order to analyze the observed peaks at temperature dependent frequencies.

In conclusion, infrared spectroscopic measurements have been performed on single crystals cobalt oxide [BiBa$_{0.66}$K$_{0.36}$O$_2$]CoO$_2$. 
The determined optical conductivity has thus exhibited a large temperature dependent transfer of spectral weight and an unconventional low energy mode.
We have shown that both its frequency and  damping scale as the temperature itself. 
So, a basic analysis has demonstrated that this mode fully scales onto a function of $\omega$/T up to room temperature. 
This behavior suggests low energy excitations of non-Fermi liquid type originating from quantum criticality in agreement with the susceptibility, the specific heat and the thermopower measurements.

\begin{acknowledgments}
We would like to acknowledge support from La R\'egion Centre, the European Union (FEDER N$^{\circ}$2620-33815) and the ANR N$^{\circ}$11-JS04-001-01.
\end{acknowledgments}


\begin{thebibliography}{20}
\bibitem{remrk}  Note that this implies fitting the reflectivity to maintain causality.
\end{thebibliography}
\end{document}